\documentclass[10pt, aps, prc, twocolumn, superscriptaddress, nofootinbib,longbibliography, showpacs, floatfix]{revtex4-2}
\pdfoutput=1
\usepackage{xcolor}
\usepackage[normalem]{ulem}
\usepackage{graphicx}   
\usepackage{dcolumn} 
\usepackage{amsmath, amsfonts}
\usepackage{amssymb}

\usepackage{orcidlink}
\usepackage{booktabs}
\usepackage{float}
\usepackage{bm}
\usepackage{hyperref}
\usepackage{multirow}

\hypersetup{breaklinks=true, colorlinks=true, linkcolor=blue, citecolor=blue, filecolor=magenta, urlcolor=cyan}

\usepackage[all]{hypcap}

\newcommand{\mean}{\phi_0}

\newcommand{\Xelements}{X_{i,k}^0}
\newcommand{\Xcelements}{X_{i,k}^c}
\newcommand{\Xb}{\bm{X^0}}
\newcommand{\Xbc}{\bm{X^c}}
\newcommand{\Sb}{\bm{S}}

\newcommand{\Ub}{\bm{U}}
\newcommand{\Vb}{\bm{V}}    

\newcommand{\ModelTotal}{m}
\newcommand{\PCTotal}{p}

\usepackage[overload]{textcase}

\begin{document}

\title{Beyond Constant Error: Heteroscedastic Bayesian Model Combination for Modeling Unmeasured Nuclei}

\author{B.~Knight\,\orcidlink{0009-0004-6777-1034}}
\email{knight@frib.msu.edu}
\affiliation{Facility for Rare Isotope Beams, Michigan State University, East Lansing, Michigan 48824, USA}
\affiliation{Department of Physics and Astronomy, Michigan State University, East Lansing, Michigan 48824, USA}

\author{S.~Lalit\,\orcidlink{0000-0001-7758-492X}}
\email{lalit@frib.msu.edu}
\affiliation{Facility for Rare Isotope Beams, Michigan State University, East Lansing, Michigan 48824, USA}

\author{P.~Giuliani\,\orcidlink{0000-0002-8145-0745}}
\email{giulianp@frib.msu.edu}
\affiliation{Facility for Rare Isotope Beams, Michigan State University, East Lansing, Michigan 48824, USA}

\author{K.~Godbey\,\orcidlink{0000-0003-0622-3646}}
\email{godbey@frib.msu.edu}
\affiliation{Facility for Rare Isotope Beams, Michigan State University, East Lansing, Michigan 48824, USA}

\author{W.~Nazarewicz\,\orcidlink{0000-0002-8084-7425}}
\email{witek@frib.msu.edu}
\affiliation{Facility for Rare Isotope Beams, Michigan State University, East Lansing, Michigan 48824, USA}
\affiliation{Department of Physics and Astronomy, Michigan State University, East Lansing, Michigan 48824, USA}

\author{A.~Ravli\'c\,\orcidlink{0000-0001-9639-5382}}
\email{ravlic@frib.msu.edu}
\affiliation{Facility for Rare Isotope Beams, Michigan State University, East Lansing, Michigan 48824, USA}
\affiliation{Department of Physics, Faculty of Science, University of Zagreb, Bijeni\v cka c. 32, 10000 Zagreb, Croatia}

\author{P.-G.~Reinhard\,\orcidlink{0000-0002-4505-1552}}
\email{Paul-Gerhard.Reinhard@fau.de}
\affiliation{Institut für Theoretische Physik, Universität Erlangen, Erlangen D-91054, Germany}

\date{\today}

\begin{abstract}
Experimentally inaccessible regions of the nuclear chart remain a challenge for global models of atomic nuclei to predict. This includes exotic nuclei near particle drip lines, superheavy elements at the extremes of mass and charge, and the neutron-rich pathways of astrophysical processes in explosive stellar environments where heavy elements are created. 
Given that individual nuclear models are imperfect, deep extrapolations are best approached using model ensembles, which allow for the systematic combination of diverse theoretical predictions. In this study, we employ the recently introduced  Bayesian Model Combination (BMC) method, based on statistical machine learning, that provides robust uncertainty quantification for forecasts using model ensembles.  To account for the inherent degradation of predictive power as models extrapolate into the yet-unexplored domain, we introduce a heteroscedastic BMC framework in which the combined theoretical uncertainty is treated as a dynamic quantity. We apply this methodology to an ensemble of realistic energy density functionals with a specific focus on the $Z=46\text{--}52$ isotopic chains. We rigorously validate the approach using both experimental data and synthetic data designed to assess performance in the deep extrapolation regime. Our results demonstrate that the proposed heteroscedastic approach yields superior calibration metrics and provides statistically principled assessments of the particle drip lines.
\end{abstract}

\maketitle

\section{Introduction}

The limits of the nuclear landscape have not been firmly established \cite{erler2012limits, Neufcourt2020b,Stroberg2021}.
The neutron-rich frontier, in particular, is challenging for nuclear theory because of the limited experimental data and the massive extrapolations involved. This part of the nuclear chart 
is of great astrophysical importance, as it hosts the rapid neutron-capture process (r-process) that is in part responsible for creating heavy nuclei~\cite{CowanRProcess}. For modeling these nuclei, nuclear density functional theory (DFT) \cite{Bender2003a,Schunck2019} has been a recognized global microscopic framework. However, many energy density functionals (EDFs) exist, and their predictions for unmeasured nuclei often disagree.

Along with the inherent systematic uncertainties associated with extrapolating far beyond the domain of data used to calibrate the functionals,  discrepancies between models introduce significant, often unquantified uncertainties into predictions of nuclear properties~\cite{Schunck2015}. This deficiency underscores the critical need for robust uncertainty quantification (UQ) methodologies to determine the predictive capabilities of nuclear DFT. In recent years, numerous machine learning techniques have been applied to nuclear mass predictions, often with the primary goal of minimizing the root mean square error (RMSE) of point predictions, see, e.g., Refs.~\cite{YuanRMSE,BNNRMSE}. While valuable, this focus can leave the reliability of the associated uncertainties underexplored.

In response to the challenge of reconciling different EDFs, recent efforts have increasingly turned to the robust framework of Bayesian statistics~\cite{Neufcourt2018, Kejzlar2023}. A landmark application of these techniques in nuclear DFT was presented in Ref.~\cite{Neufcourt2019} using Bayesian Model Averaging (BMA), a multi-model approach. The BMA systematically combines predictions from a suite of different EDFs, weighting each model based on its ability to reproduce known experimental data by itself. A known drawback of BMA is its assumption that one of the models is the ``true model"  that perfectly describes reality, a deficiency that has motivated the exploration of other model-mixing approaches~\cite{Kejzlar2023}.
In particular, Ref.~\cite{Giuliani2024} introduced 
the Bayesian Model Combination (BMC) framework that contains model preselection and model orthogonalization. This approach has recently been used to predict the proton drip lines for the tin isotopes~\cite{Ireland2026}. 

While powerful, these multi-model approaches rely on a simplifying assumption: the underlying intrinsic theoretical error can be modeled with constant (homoscedastic) variance. This assumption is not motivated by the properties of the models. The natural expectation is that a model's predictive power will degrade and its associated uncertainties will grow as it extrapolates further from the region of known data. In this work, we remove this restriction by introducing a more flexible and physically realistic statistical model. We propose that the theoretical uncertainty is not constant but dynamic (heteroscedastic), allowing it to vary across the nuclear chart. We demonstrate that such a framework provides a robust description of theoretical error and yields demonstrably superior uncertainty quantification for extrapolations of observables associated with nuclear binding energies. 

The outline of this paper is as follows. In Sec.~\ref{sec:nuc_models}, we describe the nuclear DFT models used in this work. The BMC methodology is discussed in Sec.~\ref{sec:bmc_method}. Section~\ref{sec:Extrapolation} describes the validation of BMC on the synthetic data. Predictions trained on experimental data are presented in  Sec.~\ref{sec:valid}. Finally in Sec.~\ref{sec:conc} we present our conclusions and outlook.

\section{Nuclear models and binding energy indicators}
\label{sec:nuc_models}
All the theoretical models we use in this work are based on nuclear DFT, with varying EDFs and calibration schemes. In particular, we use three covariant EDFs: DD-PC1 \cite{Niksic2008}, DD-PCX \cite{Yuksel2019} and DD-ME2 \cite{Lalazissis2005}, two Skyrme EDFs: UNEDF1 \cite{Kortelainen2012a} and UNEDF2 \cite{Kortelainen2014a}, as well as one Fayans EDF: Fy(IVP) \cite{Karthein2024}. The covariant EDFs employ the separable pairing functional as described in Ref.~\cite{Ma2009}, the Skyrme-based EDFs have mixed contact delta pairing~\cite{Kortelainen2012a}, while the Fayans EDF has more sophisticated pairing terms including the gradient terms~\cite{Reinhard2017a}. The underlying Hartree-Fock-Bogoliubov (or Relativistic Hartree-Bogoliubov in case of covariant EDFs) equations are solved either by expanding the quasiparticle states in the basis of an axially-deformed harmonic oscillator~\cite{Niksic2014a,Marevic2022a} or discretizing them in the coordinate space~\cite{Reinhard2021a}. All calculations assume axially-deformed and reflection-invariant nuclear shapes. Calculations of odd-$A$ and odd-odd nuclei are performed by self-consistent quasiparticle blocking assuming equal-filling approximation~\cite{Perez-Martin2008a}. 

To establish a rigorous framework for our evaluation, we partition the statistical domain of our models into distinct datasets for calibration and testing. Specifically, the dataset is divided into three distinct subsets:
\begin{itemize}
    \item{\textbf{Training Set}: This subset comprises the core of the experimentally known domain from the Atomic Mass Evaluation (AME2020), localized near the valley of stability. It serves as the calibration dataset to determine both the BMC model weights and the coefficients of the error model.}
    \item{\textbf{Validation Set}: To evaluate performance under extrapolation, we define a validation set isolated at the outermost boundaries of the experimentally known domain. This boundary region consists of the lightest and heaviest $n$ measured nuclei along each isotopic chain, where $n$ is dynamically chosen to guarantee a 30/70 validation-to-training data split within the known nuclear chart. The explicit role of this validation data is to evaluate the performance of each error model against each other.}
    \item{\textbf{Prediction (Extrapolation) Set}: This subset is made up of every nucleus where theoretical model predictions exist but experimental data do not, serving as our deep extrapolation domain.}
\end{itemize}

The observables we focus on in this work are a set of binding energy indicators, derived from the total binding energy $BE(Z,N)$ of a nucleus with $Z$ protons and $N$ neutrons. In particular, these are the one- and two-neutron, and one- and two-proton separation energies, defined as the finite differences in binding energy between neighboring isotopes and isotones, respectively:
\begin{subequations}\label{eq:Sep}
\begin{align}
S_{1n}(Z,N) &= BE(Z,N) - BE(Z,N-1), \label{eq:Sep_a}\\
S_{2n}(Z,N) &= BE(Z,N) - BE(Z,N-2), \label{eq:Sep_b}\\
S_{1p}(Z,N) &= BE(Z,N) - BE(Z-1,N), \label{eq:Sep_c}\\
S_{2p}(Z,N) &= BE(Z,N) - BE(Z-2,N). \label{eq:Sep_d}
\end{align}
\end{subequations}
We also calculate the alpha decay Q-values:
\begin{align}
Q_\alpha(Z,N) &= BE(Z-2,N-2) + BE(2, 2) \notag \\
&\quad - BE(Z,N). \label{eq:Qa}
\end{align}

\begin{figure}[hbtp]
    \includegraphics[width=\linewidth]{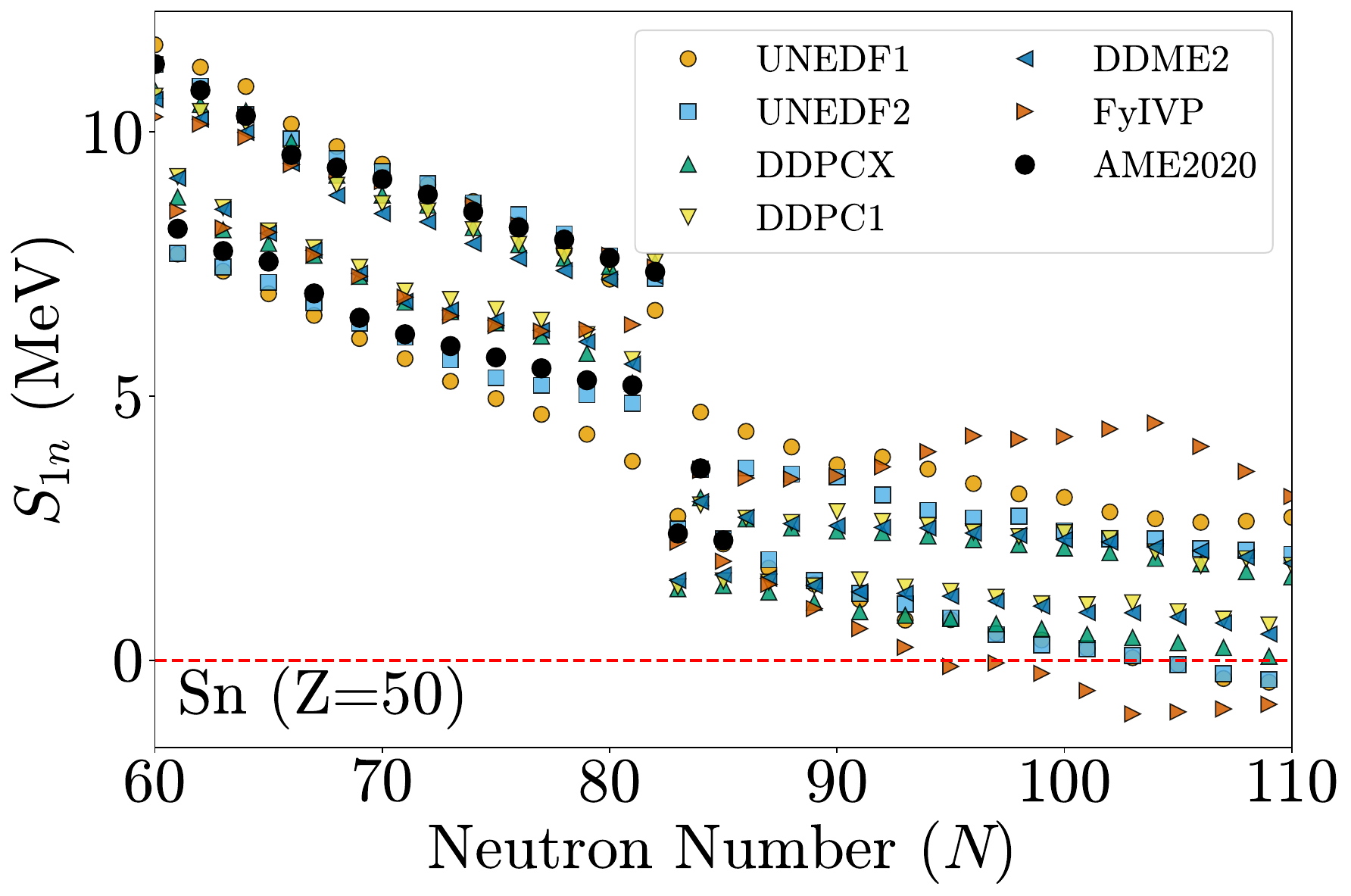}
    \caption{Predictions of $S_{1n}$ for the Sn ($Z=50$) isotopic chain from the six DFT models, compared to experimental AME2020 data~\citep{AME2020}. In the experimentally known region ($N \lesssim 82$), the models show reasonable agreement with data, but their spread grows substantially as they extrapolate toward the neutron drip line. The drip line threshold ($S_{1n}=0$) is marked in red.}
    \label{fig:s1n_models}
\end{figure}

While it is possible to train the BMC model entirely on $BE(Z,N)$ and derive the separation energies from the predicted total binding energies, this method fails to account for the local correlations between adjacent nuclei in the isotopic chain. To ensure the most accurate predictions possible, we perform calibrations on each separation energy indicator separately. This yields significantly tighter uncertainty bounds across the training and extrapolation datasets. The average predicted uncertainty for $S_{1n}$ is 0.35 MeV, compared to $1.12$ MeV when propagating errors from a $BE(Z,N)$ calibration using the same error model. 

Figure~\ref{fig:s1n_models} displays the $S_{1n}$ predictions of the six models for the tin chain. The models track the experimental separation energies well in the region calibrated by AME2020 data~\citep{AME2020}, but diverge beyond the last measured isotope $^{135}$Sn. 
This growing inter-model spread signals increasing theoretical uncertainty across the collection of models, and provides one of the key quantities we use to estimate the error scale we introduce in Sec.~\ref{sec:bmc_method}.

\section{Bayesian Model Combination}
\label{sec:bmc_method}
 
Our application of BMC closely follows Ref.~\citep{Giuliani2024}, where the models discussed in the previous section satisfy the ``reasonable model" criterion defined therein.
In the following, we present a brief overview of the method and highlight new developments, particularly those concerning the heteroscedastic likelihood model.

To construct the BMC, a set of $m$ existing models $\mathcal{M}_k, k \in \{1, 2, \ldots, m\}$, with $n$ predictions for observables $\boldsymbol{y}\equiv (y_i)$, $i \in \{1, 2, \ldots, n\}$, is carefully combined through an experimental dataset.
In contrast to other model combination approaches~\citep{Neufcourt2019, Kejzlar2023}, the BMC method combines the dominant principal components (PCs) of the model ensemble, aiming to enhance predictive capability by mitigating overfitting while reducing the influence of redundant or closely related models on the final mixture.

Following Ref.~\cite{Giuliani2024} we construct a matrix of model calculations $\Xb\equiv (\Xelements)$, where each column $k$ represents the predictions of model $\mathcal{M}_k$ across the observables of interest $y_i$ (binding energies, separation energies, etc).
The centered matrix  of model calculations $\Xbc \equiv (\Xcelements)$
is then given by
\begin{equation}
\Xcelements =  \Xelements - \mean(x_i),
\end{equation}
where 
\begin{equation}
    \mean(x_i)\equiv\frac{1}{m}\sum_{k=1}^m \mathcal{M}_k(x_i)
\end{equation}
is the average model prediction vector and $\boldsymbol{x}\equiv(x_i)$ the vector  of inputs   defining observable $y_i$ such as proton $Z_i$ and neutron $N_i$ numbers, as well as the observable class such as the separation energy indicators in Eqs.~\eqref{eq:Sep}. Each column $k$ of the centered matrix therefore characterizes how a model $\mathcal{M}_k$ deviates from the average model prediction $\mean$. We decompose this matrix using the singular value decomposition (SVD) algorithm~\cite{blum_hopcroft_kannan_2020}:
\begin{align}\label{Eq: SVD}
    \Xbc_{{n\times \ModelTotal}} &= \Ub_{{n\times n}} \ \  \Sb_{{n \times \ModelTotal}} \ \  \Vb^T_{{\ModelTotal \times \ModelTotal}} \notag \\
    &\approx \ \ \hat \Ub_{{n\times  \PCTotal}} \ \  \hat \Sb_{{ \PCTotal \times  \PCTotal}} \ \  \hat \Vb^T_{{ \PCTotal \times  \ModelTotal}},
\end{align}
and create an approximate representation of the centered matrix ${\bm{\hat{X}}}^c_{{n\times \ModelTotal}}$ by truncating its dimensions and keeping the first $p$ singular vectors.
This truncation is a dimensionality reduction approach that provides the principal components~\cite{jolliffe2002principal}, which represent the axes of maximum variability from the mean $\mean$ among the collection of models. These components, denoted by $\phi_j(\boldsymbol{x})$ with $j \in \{1, 2, \ldots, p\leq m\}$, are the $j$-th retained principal components obtained from the SVD of the centered model matrix $X^c$. These form the basis vectors for a lower-dimensional PC space.  The subset of components retained for the BMC is represented by this basis.
The number of retained components $p$ is a hyperparameter of the model, and is usually selected by evaluating the performance on a set of validation data, similar to traditional machine learning approaches~\cite{brunton2019data}.

For a chosen value $p$, the BMC is written as:
\begin{equation}\label{Eq: MC}
f^{\dagger}(x_i;\boldsymbol{b}) = \phi_0(x_i) + \sum_{j=1}^{p}b_{j}\phi_{j}(x_i),
\end{equation}
where the weights $\boldsymbol{b}\equiv (b_j)$ represent the free parameters of the BMC model and are calibrated within a Bayesian framework using the selected experimental data. The likelihood model we use expresses the relationship between observations $y_i$ and BMC predictions  as:
\begin{equation}\label{Eq: relationship}
    y_i = f^{\dagger}(x_i;\boldsymbol{b}) + \varepsilon_i \sigma_i,\quad \varepsilon_i \sim \mathcal{N}(0, 1),
\end{equation}
where the notation $\varepsilon_i \sim \mathcal{N}(0, 1)$  indicates that the random variable $\varepsilon_i$ is drawn from a normal (Gaussian) distribution with mean 0 and variance 1. We represent the error, or discrepancy, between model predictions $f^{\dagger}(x_i;\boldsymbol{b}) $ and experimental data $y_i$ through the term $\varepsilon_i\sigma_i$, with $\boldsymbol{\sigma}\equiv(\sigma_i)$ being the collection of error scales for all data. 

In Ref.~\cite{Giuliani2024}, the authors modeled the discrepancies in binding energies to be statistically independent of each other. In this work, we focus on the separation energies defined in Eqs.~\eqref{eq:Sep}. Since these are defined as differences between binding energies, we treat their corresponding discrepancies as also statistically independent.
For the size of the errors, possibly the simplest assumption is that they share the same scale $\sigma_i\equiv \sigma_0$, and this overall value becomes a variable to be estimated through the Bayesian framework~\cite{Gelman2004d}. 
We refer to this approach as the \textit{homoscedastic error model} (HoEM), which is the same procedure adopted in Refs.~\cite{Giuliani2024,Ireland2026}. 

One particular weakness with the homoscedastic variance assumption when extrapolating is the highly discrepant behavior of different functionals outside of the region constrained by experimental observation. As illustrated in Fig.~\ref{fig:s1n_models}, while each model is well-constrained in the region near stability where training data exist, the differing physical assumptions among the models, as well as variations in their calibration datasets, drive the models' predictions apart as they approach the extrapolation regime. In regions of significant model disagreement, a homoscedastic assumption artificially enforces a constant uncertainty bandwidth, underestimating the quality of fit in regions of the training domain that the ensemble reproduces well, while overestimating it in the predictive domain. 

An alternative approach is to allow the error scale to vary across data points, a more flexible method that can improve the robustness of the predictions~\cite{carroll2017transformation}.  Since in this study we are interested in the ability to extrapolate beyond the regions of experimental data, we need to construct a function that estimates how the error scale varies across all isotopes~\cite{davidian1987variance}:
\begin{equation}\label{Eq: sigma}
   \sigma_i \equiv g(\boldsymbol{z}_i;\Lambda).
\end{equation}

The error scale $\sigma_i$ depends on each data point through  $\boldsymbol{z}_i$, which are a set of features - or predictors - that, as we explain in more detail later, can themselves depend on the inputs $x_i$, and we expect to be correlated with the actual model error. The parameters $\Lambda$  are learned within a Bayesian framework, and help characterize the relationship between $\sigma_i$ and the features $\boldsymbol{z}_i$. We refer to this approach as the \textit{heteroscedastic error model} (HeEM).

Once we have specified the model combination $f^\dagger$ in Eq.~\eqref{Eq: MC}, its relationship with the experimental observations in Eq.~\eqref{Eq: relationship}, and the expression for the error scale $\sigma_i$ in Eq.~\eqref{Eq: sigma}, we construct our likelihood model~\cite{Gelman2004d} assuming conditional independence on the observations as:
\begin{equation}\label{Eq: likelihood}
p(\boldsymbol{y}|\boldsymbol{b},\Lambda) \propto \prod_{i=1}^{n} \frac{1}{\sigma_i} \exp\left(-\frac{(f^{\dagger}(x_i;\boldsymbol{b})-y_i)^2}{2\sigma_i^2}\right).
\end{equation}
We note that the HoEM approach is contained within the more general HeEM approach by setting $g(z_i;\Lambda) = \sigma_0$ in Eq.~\eqref{Eq: sigma}, with $\Lambda=\sigma_0$ the only parameter to be estimated. By combining the constructed likelihood with a suitable prior $p(\boldsymbol{b},\Lambda)$, we construct the posterior distribution that will allow us to constrain the parameters of the BMC given the available experimental data:
\begin{equation}\label{Eq: posterior}
p(\boldsymbol{b}, \Lambda | \boldsymbol{y}) \propto p( \boldsymbol{y} | \boldsymbol{b},  \Lambda ) p(\boldsymbol{b},  \Lambda ).
\end{equation}
For the prior distribution of the weights $\boldsymbol{b}$, we follow the same approach as in Ref.~\cite{Giuliani2024}, creating an independent normal distribution centered at zero with widths informed by the SVD decomposition. The rationale behind this choice is that we believe the true experimental measurements to be relatively close to the span of reasonable models $\mathcal{M}_k$, and as such we expect the BMC $f^\dagger$ to follow a similar power spectrum in the coefficients $\boldsymbol{b}$ as the models themselves. For the HoEM we also follow the approach in Ref.~\cite{Giuliani2024} with a weakly informed prior through a Gamma distribution for the precision $1/\sigma_0^2$.  In all cases, samples from the posterior, Eq.~\eqref{Eq: posterior}, are obtained by using a Gibbs sampler. For the homoscedastic model this is a standard two-block Gibbs sampler, while the heteroscedastic model uses Metropolis–Hastings step embedded within each Gibbs sweep (a Metropolis-within-Gibbs scheme)~\cite{hoff2009f}. The point predictions reported in our validation tables and figures represent the posterior expectation values obtained by averaging over the full MCMC chain generated by the Gibbs sampler. Because the BMC framework is linear in its weights, this posterior predictive mean corresponds to evaluating the model at the center of mass of the posterior distribution,  preserving full alignment with our sampled uncertainty calibration metrics.

The uncertainties we report alongside these point predictions are constructed in an analogous manner. Rather than evaluating the error-model expression in Eq.~\eqref{Eq: sigma} at a single point estimate of $\Lambda$, we propagate every retained sample of $(\boldsymbol{b},\Lambda)$ from the Gibbs chain through the model $f^\dagger$, generating a set of posterior predictions for each nucleus and observable. The reported uncertainty is then taken to be the 68\% credible interval (16th–84th percentile) of the set of sampled predictions, rather than an analytic propagation of a single fitted $\sigma_i$. This procedure ensures that the quoted uncertainty reflects both the spread in the BMC weights $\boldsymbol{b}$ and the spread in the error-model parameters $\Lambda$.


\subsection{Heteroscedastic Error Model Design}
\label{sec:results}

 In order to develop an error model that describes the confidence of the BMC, we must identify statistical indicators that are correlated with the model's performance. One primary driver of model performance degradation is calibration drift: a statistical framework is typically well-calibrated only where training data are available. Consequently, the distance of a given nucleus from the training domain of the Bayesian calibration emerges as a natural indicator for error scaling.
 
\begin{figure}
    \centering
    \includegraphics[width=1\linewidth]{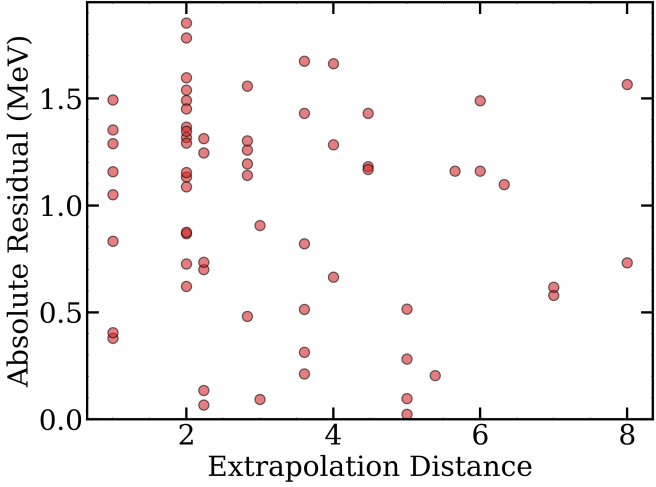}
    \caption{For each nucleus in the validation set, plotted is the  distance (\ref{T-dist}) and the residual between the mean homoscedastic prediction and the experimental value.}
    \label{fig:extrapexample}
\end{figure}

  An intuitive space for evaluating this proximity is the $(N,Z)$ grid of the nuclear domain, where the  distance to the nearest training isotope is
  \begin{equation}\label{T-dist}
    \Delta(N, Z) = \min_{(N_i, Z_i) \in \mathcal{T}} \sqrt{(N - N_i)^2 + (Z - Z_i)^2},
    \end{equation}
    where $\mathcal{T}$ is the  training set.
  However, evaluating distance over a discrete integer lattice severely restricts the granularity and fidelity of the metric. Empirically, this coordinate distance exhibits a weak correlation with the observed residuals, rendering it an ineffective predictor of model confidence (see Fig.~\ref{fig:extrapexample}). 

\begin{figure}[!htbp]
    \centering
    \includegraphics[width=1\linewidth]{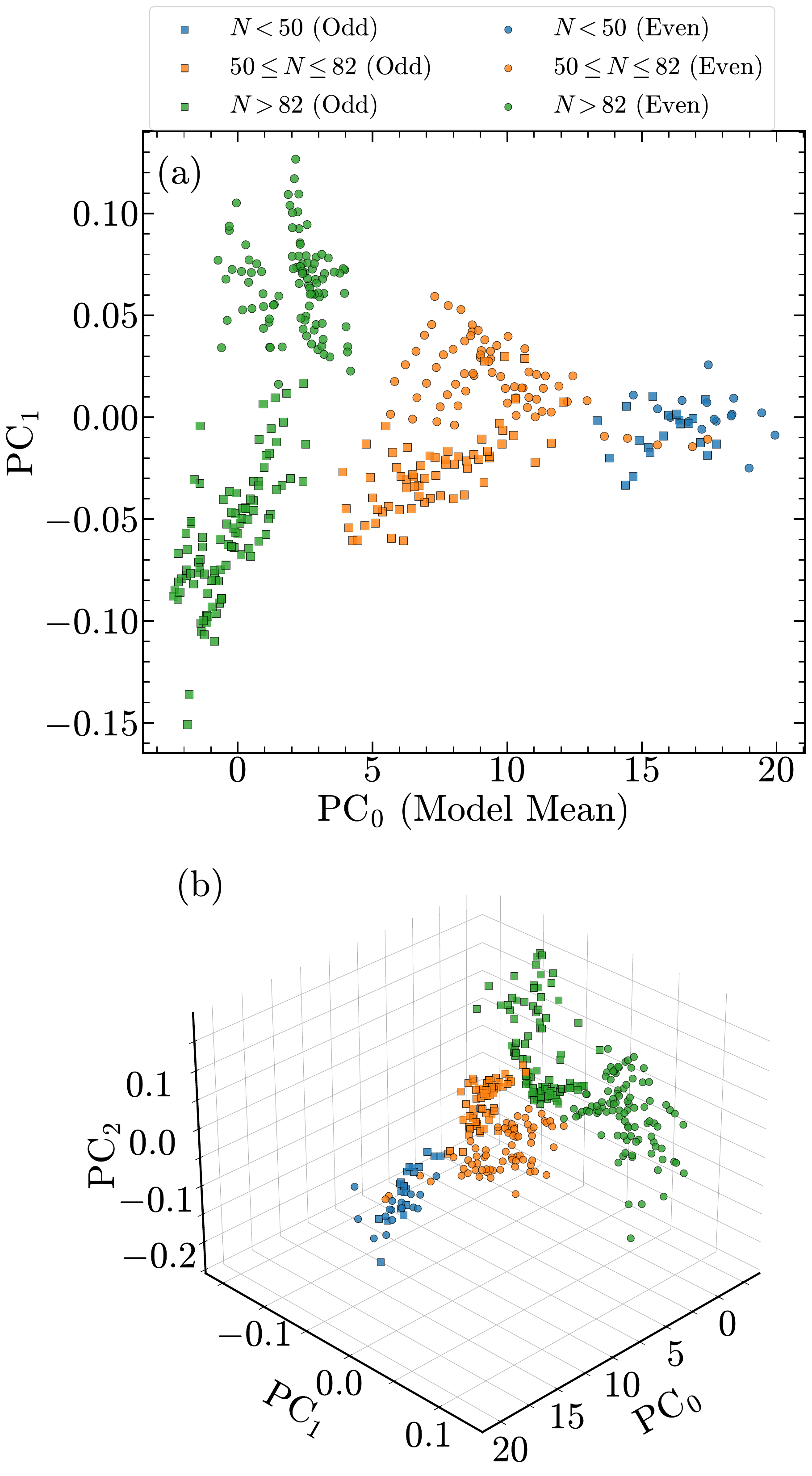}
    \caption{2D and 3D visualizations of the model data in PC-space. The grouping of nuclei of the same neutron shell and parity shows how nuclear structure is preserved in the principal component space.}
    \label{fig:combined-plot}
\end{figure}

To obtain a more physically informative metric, we may instead assess nuclei in our predictive domain within the continuous space spanned by the dominant PCs. Projecting the model predictions onto the primary axes of variability maps each nucleus into a continuous, lower-dimensional PC space. For a given nucleus $i$, we construct a centered deviation vector $\mathbf{x}_i^c = \mathbf{x}_i^0 - \phi_0(x_i)$, where $\mathbf{x}_i^0$ represents the raw ensemble predictions and $\phi_0(x_i)$ is the average model prediction vector. The coordinates $\mathbf{\zeta}_i$ of the nucleus in the reduced $p$-dimensional PC space are then obtained by projecting this deviation vector onto the truncated right singular vectors $V_{m \times p}$ determined from the SVD of the training matrix:
\begin{equation}
\mathbf{\zeta}_i = \mathbf{x}_i^c V_{m \times p}
\end{equation}
where $V_{m \times p}$ contains the first $p$ columns of $V$, defining the orthogonal axes of maximum variability across the model ensemble. As shown in Fig.~\ref{fig:combined-plot}, the nuclei in the PC space are clustered by familiar nuclear properties such as nuclear shell closure and odd-even parity. This emergent structure demonstrates that the principal components capture systematic trends in the model predictions. Principal Component Analysis (PCA) naturally isolates the physical variations of the nuclei into orthogonal principal components, resulting in a continuous PC space that acts as a physical spectrum, organizing nuclei automatically by their structural properties.
The continuous nature of this space ensures sufficient resolvable fidelity in the uncertainty predictor to accurately quantify a model's distance from the training data.

A second driver of performance degradation we consider is a larger inter-model spread, where the individual models exhibit significant mutual disagreement. 
In particular, a high model spread for a specific nucleus could indicate disagreement in the underlying physics model assumptions. This spread can be quantified as the variance $\sigma^2$ of the individual model predictions for each nucleus.

These two numerical predictors are taken as quantifiers of the primary mechanisms of degradation in prediction quality:
\begin{itemize}
    \item Distance $d_i$ from the center of the training data in PC space;
    \item  Variance $v_i$ of model predictions. 
\end{itemize}

\begin{figure}[!htbp]
    \centering
    \includegraphics[width=1\linewidth]{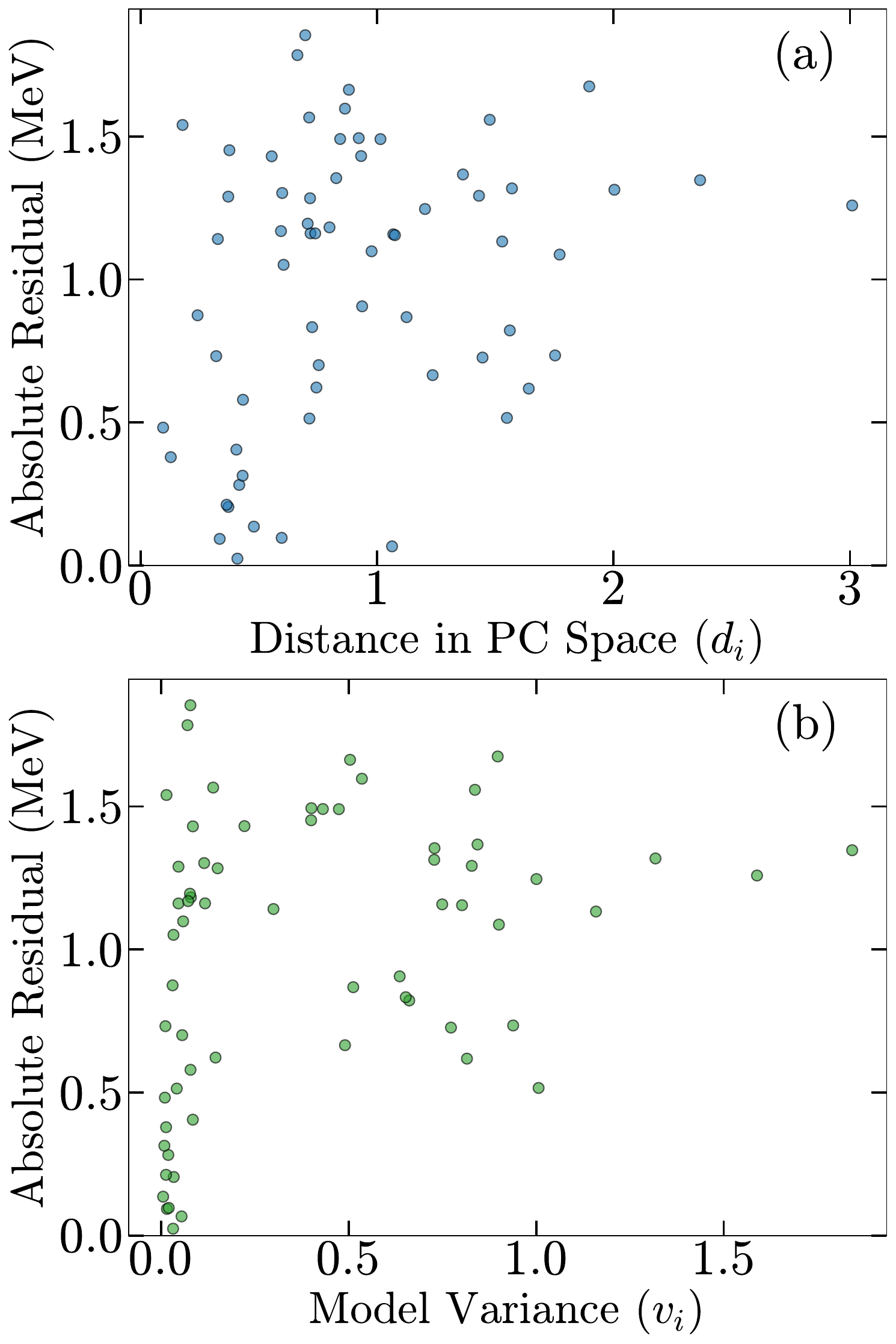}
    \caption{Two uncertainty predictors plotted against the residual between the mean homoscedastic prediction and the true experimental values. Training data are omitted to avoid cluttering the plot.}
    \label{fig:predictors}
\end{figure}

Figure~\ref{fig:predictors} shows that the cross-variation between the two defined predictors and the absolute residual of the constant-variance BMC indeed share some level of correlation. Using these two predictors, we can write the error models we explore in this work:
\begin{subequations}\label{eq:forms}
\begin{eqnarray}
   \text{(HoEM) } \sigma_i^2 &=& \alpha, \label{eq:eqHoEM} \\
   \text{(HeEM Linear) } \sigma_i^2 &=& \alpha + \beta_1 d_i + \gamma_1 v_i, \label{eq:eqHeEMLin} \\
   \text{(HeEM Quadratic) } \sigma_i^2 &=& \alpha + \beta_{1} d_i + \beta_{2} d_i^2 \nonumber \\
    && + \gamma_{1} v_i + \gamma_{2} v_i^2 \nonumber \\
    && +\delta d_i v_i, \label{eq:eqHeEMQuad}
\end{eqnarray}
\end{subequations}
where $\alpha$, $\beta_j$, $\gamma_j$, and $\delta$ are the error model parameters $\Lambda$ sampled from the constructed posterior distribution given by Eq.~(\ref{Eq: posterior}). Positivity of $\sigma_i^2$ is enforced by restricting $\alpha,\beta_j,\gamma_j,\delta > 0$ through their priors and by rejecting any Metropolis--Hastings proposal that violates this constraint. The first expression, Eq.~(\ref{eq:eqHoEM}), corresponds to the constant-variance homoscedastic error model.

\subsection{Evaluation Metrics}
To evaluate these error models rigorously, we introduce several statistical metrics to quantify the quality of the uncertainty calibration. RMSE is a metric that evaluates the average magnitude of the residual between a model's predictions and the target values. While RMSE is capable of evaluating the performance of the mean predictions of a model, it is not sensitive to the uncertainty of a model, and therefore not useful in describing the performance of an error model. 

A better suited metric that takes into account the model uncertainty is the reduced $\chi^2$, calculated as:
\begin{equation}
\chi^2 = \frac{1}{N} \sum_{i=1}^{N} \frac{(y_{\text{pred},i} - y_{\text{exp},i})^2}{\sigma_i^2},
\label{eq:chi2}
\end{equation}
where $N$ is the total number of data points. 
While the $\chi^2$ score is an accurate estimator of whether a model's variance is globally well-scaled, it is not a complete measure of the distribution's quality. For example, if a model is generally underconfident but exhibits extreme overconfidence in specific cases, these opposing deviations may aggregate to a misleading total error. To obtain a more robust metric that accounts for these conflicting deviations, we analyze the empirical coverage across various credible intervals (CI). A posterior predictive distribution for a dataset is compared with the percentage of true data points that fall within a given nominal credible interval $p$. 

To aggregate the deviation of the empirical coverage ($C_\text{emp}$)~\cite{Schall2012,Morris2019} from the nominal credible interval ($p$) across the full probability distribution, we compute the \textit{mean absolute calibration error} (MACE):
\begin{equation}
   \text{MACE} = \frac{1}{M} \sum_{j=1}^\text{M}|C_\text{emp}(p_j) - p_j|,
\label{eq:mace}
\end{equation}
where $M$ is the number of samples in the validation or test dataset.
A lower MACE value indicates a model that is better calibrated, meaning its predicted probabilities more accurately reflect the frequency of observed outcomes. Global metrics across credible intervals generalize the concept of predictive interval calibration to continuous regression tasks~\cite{Kuleshov2018}.

\section{Validation of the BMC on synthetic data}
\label{sec:Extrapolation}

Having defined the various BMC error models and our evaluation metrics, in this section we proceed to test their extrapolation performance in a synthetic data experiment.
In this procedure, each constituent DFT model is treated as a surrogate for the 
experimental data. 
We construct the BMC framework using the remaining set of models and calibrate it through synthetic data constrained exclusively to the experimentally known domain (AME2020). This calibration data consists of the numerical predictions generated by the DFT model that was excluded from the ensemble. By analyzing the performance of the BMC on the holdout test data set, we can compare how adopting the various error models impacts the calculated observables well outside the training domain, for example, near the driplines.

\begin{figure}[htb]
    \includegraphics[width=1\linewidth]{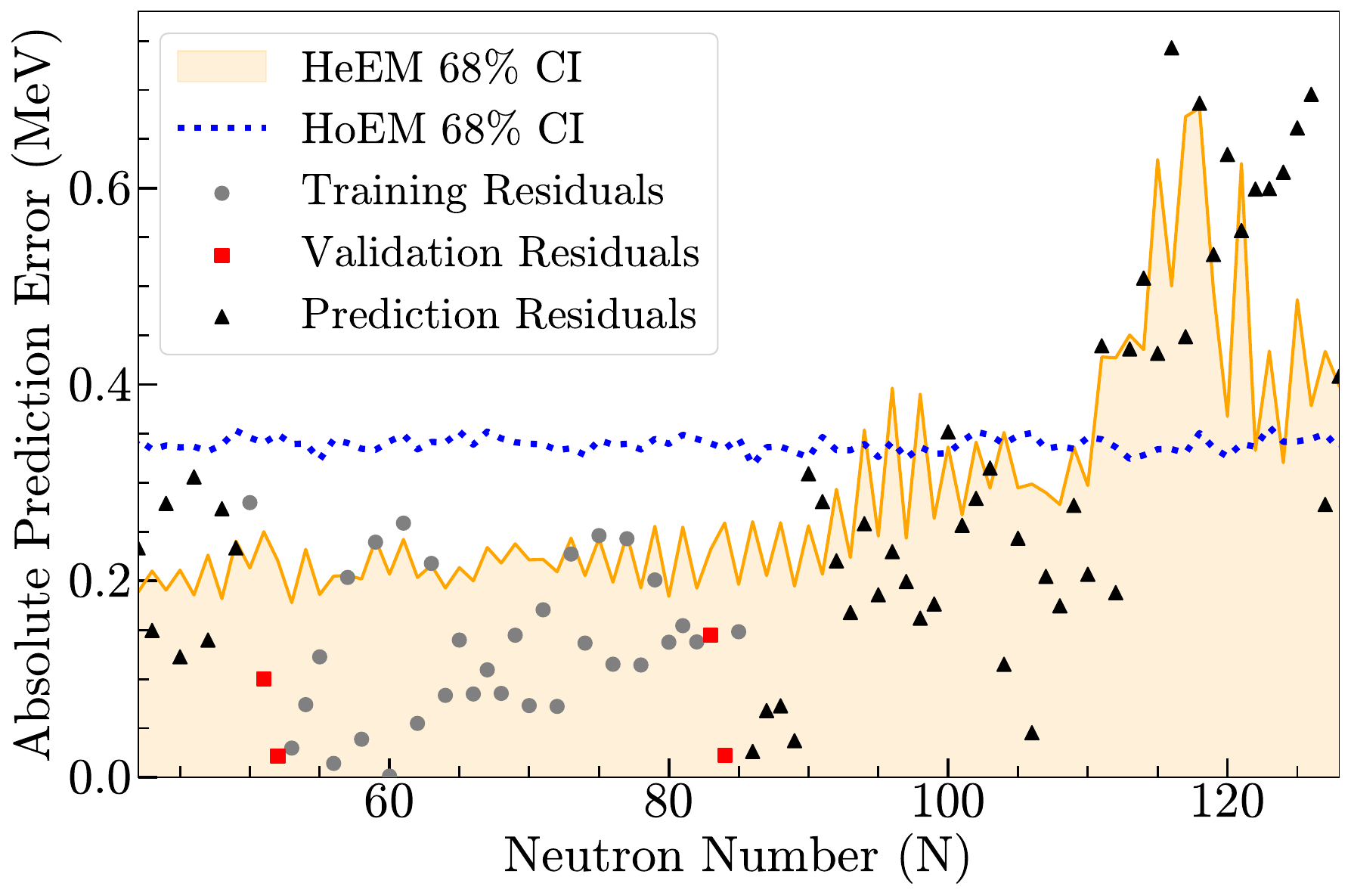}
    \caption{Visualization of the synthetic data test. In this case, $S_{1n}$  is predicted for the tin chain using  DD-PC1 as the synthetic data model. The linear HeEM is used.}
    \label{fig:syntheticdataexperiment}
\end{figure}

Figure~\ref{fig:syntheticdataexperiment} visualizes the synthetic validation for the case where the synthetic data are produced by DD-PC1. The homoscedastic model retains a static error scale determined by the training data, leading to underconfident predictions in the training regime and overconfident predictions in the extrapolation regime. In contrast, the linear heteroscedastic error model allows the credible interval to expand and successfully encompass the synthetic data even as it deviates significantly from the predicted mean.

Table \ref{tab:combined_metrics} shows the RMSE and reduced $\chi^2$ results of the homoscedastic and heteroscedastic error models compared against each other. For four of the six synthetic data models tested, the HoEM deviates significantly from unity. However, in almost all of these cases, both the linear and quadratic heteroscedastic models are closer to unity. This indicates that allowing the error magnitude to vary with identified indicators can address and rectify both systematically overconfident and systematically underconfident error models. 

\begin{table}[ht!]
\caption{RMSE and reduced $\chi^2$ values for both error models predicting the $S_{1n}$ values for the holdout validation set. Lower RMSE is better, and $\chi^2$ closer to unity is better.}
\label{tab:combined_metrics}
    \begin{ruledtabular}
\begin{tabular}{l ccc c ccc}
 & \multicolumn{3}{c}{RMSE (MeV)} & & \multicolumn{3}{c}{Reduced $\chi^2$} \\
 Synthetic & & Lin. & Quad. & & & Lin. & Quad. \\
data & HoEM & HeEM & HeEM & & HoEM & HeEM & HeEM \\
\hline\\[-7pt]
DDME2   & 0.38 & 0.37 & 0.37 & & 1.70 & 1.29 & 1.31 \\
DDPC1   & 0.22 & 0.22 & 0.22 & & 0.46 & 0.71 & 0.74 \\
DDPCX   & 0.42 & 0.42 & 0.42 & & 2.45 & 1.43 & 1.52 \\
Fy(IVP)    & 1.62 & 1.64 & 1.65 & & 1.30 & 1.16 & 1.26 \\
UNEDF1  & 0.40 & 0.40 & 0.39 & & 1.05 & 0.92 & 0.88 \\
UNEDF2  & 0.29 & 0.30 & 0.30 & & 0.95 & 0.94 & 1.00 \\[4pt]
Average & 0.55 & 0.56 & 0.56 & & 1.32 & 1.08 & 1.12 \\
\end{tabular}
    \end{ruledtabular}
\end{table}

However, the improvement in $\chi^2$ does not definitively demonstrate a more accurate error model. To do so, the MACE scores are calculated sequentially, treating each DFT model as the source of synthetic data. The resulting MACE scores were averaged to establish comprehensive benchmarks for the synthetic data test. The results for $S_{1n}$ are listed in Table~\ref{tab:calibration_scores_s1n}.
We find that both the linear and quadratic HeEMs lower the calibration error compared to the HoEM baseline within the validation boundaries. Crucially, even as model performance degrades within the extrapolation domain, metrics on the validation set remain highly predictive of performance across both prediction sets considered.

\begin{table}[ht!]
\caption{MACE score averaged across error models predicting synthetic $S_{1n}$. The performance is given for two prediction sets: Prediction (Nearby) includes the eight closest nuclei on either side of the known data along each isotopic chain, and Prediction (All) encompasses the full predictive domain. Lower MACE is better.}
\label{tab:calibration_scores_s1n}
    \begin{ruledtabular}
\begin{tabular}{l cc c cc}
 & & & & \multicolumn{2}{c}{Prediction} \\
Error model & Train & Validation & & (Nearby) & (All) \\
\hline\\[-7pt]
HoEM             & 5.88 & 10.90 & & 11.66 & 11.10 \\
HeEM (Linear)    & 4.15 &  7.31 & &  7.61 &  8.26 \\
HeEM (Quadratic) & 4.56 &  7.21 & &  8.02 &  8.44 \\
\end{tabular}
    \end{ruledtabular}
\end{table}

The theoretical models share a microscopic framework that introduces systematic biases, and consequently, the entire ensemble may collectively deviate from reality. Therefore, the true experimental values may be more challenging to predict than the synthetic calibration data. Despite this limitation, the synthetic experiment shows that the HeEM's calibration is not just an artifact of a single benchmark. Averaged over all six synthetic tests, the linear HeEM's validation MACE (7.31) is close to its prediction-set MACE (8.26). Overall, validation-set metrics are a reasonable proxy for extrapolation accuracy, at least within the scope of the DFT models used for this calibration.

Taken together, these synthetic benchmarks inform our choice of error model for the analysis of the real experimental data. While the quadratic HeEM achieves marginally tighter calibration on the validation set in some cases (see Table~\ref{tab:calibration_scores_s1n}), this advantage does not consistently carry over to the prediction sets, where the linear HeEM matches or outperforms its quadratic counterpart. Given that the linear HeEM achieves this performance with three error-model parameters compared to six for the quadratic form, we consider the linear form to be the more robust choice for extrapolation. We therefore adopt the linear HeEM as our primary heteroscedastic model when applying the BMC framework to the experimental AME2020 dataset in Sec.~\ref{sec:valid}.

\section{Model Application to AME2020 dataset}
\label{sec:valid}
Having validated the framework using synthetic data, we now apply BMC to extrapolate from the AME2020 dataset in the region of $Z=46$ to $Z=52$.
For the training data, we use the binding energies, $Q_\alpha$ values, and neutron separation energies from $N=40$ to $N=130$ and $Z=46,48,50,52$. We use proton separation energies for $Z=48$ to $Z=52$ and from $N=40$ to $N=80$.

\begin{table}[ht!]
\caption{RMSE results on the holdout validation set for separation energies (in MeV) for the BMC models and the individual models used to train BMC.}
\label{tab:rmse_sep}
    \begin{ruledtabular}
\begin{tabular}{l cc c cc}
 & \multicolumn{2}{c}{Neutron (MeV)} & & \multicolumn{2}{c}{Proton (MeV)} \\
Model & $S_{1n}$ & $S_{2n}$ & & $S_{1p}$ & $S_{2p}$ \\
\hline\\[-7pt]
BMC (Homoscedastic)   & 0.22 & 0.27 & & 0.26 & 0.36 \\
BMC (Linear HeEM)     & 0.22 & 0.27 & & 0.27 & 0.37 \\
BMC (Quadratic HeEM)  & 0.22 & 0.28 & & 0.31 & 0.40 \\[4pt]
DDME2                 & 0.80 & 0.70 & & 0.81 & 0.93 \\
DDPC1                 & 0.80 & 0.81 & & 0.76 & 0.80 \\
DDPCX                 & 0.66 & 0.75 & & 0.79 & 1.12 \\
Fy(IVP)               & 0.58 & 0.70 & & 1.26 & 1.12 \\
UNEDF1                & 0.43 & 0.53 & & 1.05 & 0.84 \\
UNEDF2                & 0.35 & 0.44 & & 0.90 & 0.88 \\
\end{tabular}
    \end{ruledtabular}
\end{table}

Table~\ref{tab:rmse_sep} shows the RMSE of the separation energies for the validation set using homoscedastic and heteroscedastic error models. For both HoEM and HeEM, the BMC process yields RMSE performance that far exceeds the performance of any individual microscopic model. This, again, demonstrates BMC's core capability to predict nuclear observables with greater accuracy than its constituent models.

\begin{table}[ht!]
\caption{Reduced $\chi^2$ values for BMC models on the holdout validation set. $\chi^2$ closer to unity is better.}
\label{tab:chi2_sep}
    \begin{ruledtabular}
\begin{tabular}{l cc c cc}
 & \multicolumn{2}{c}{Neutron} & & \multicolumn{2}{c}{Proton} \\
Model & $S_{1n}$ & $S_{2n}$ & & $S_{1p}$ & $S_{2p}$ \\
\hline\\[-7pt]
HoEM            & 0.57 & 0.66 & & 0.49 & 0.69 \\
Linear HeEM     & 0.86 & 0.86 & & 0.71 & 0.74 \\
Quadratic HeEM  & 0.81 & 0.75 & & 0.55 & 0.53 \\
\end{tabular}
    \end{ruledtabular}
\end{table}

As shown in Table \ref{tab:chi2_sep}, the HoEM consistently produces $\chi^2$ scores significantly below unity. This indicates that the global variance assumption forces the model to adopt overly conservative uncertainties on the validation set to accommodate outliers. In contrast, the HeEM formulations achieve values closer to unity, demonstrating superior calibration of the error scale. Additionally, the linear form of the error model outperforms the quadratic form, especially for proton separation energies.

\begin{figure}[!htbp]
    \centering
    \includegraphics[width=1\linewidth]{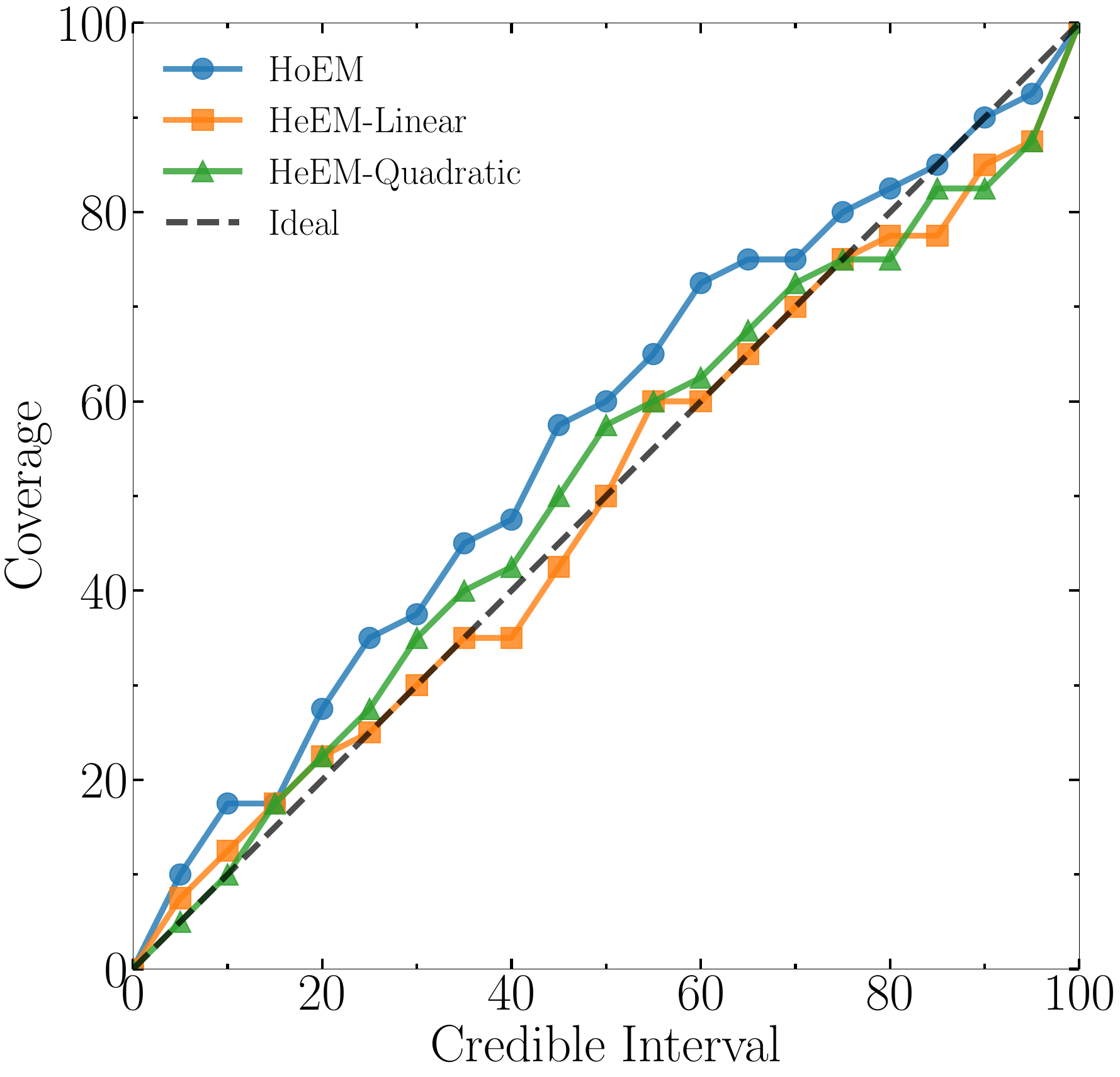}
    \caption{Empirical coverage probability  for different models of $S_{1n}$.} 
    \label{fig:coverage}
\end{figure}

In Fig.~\ref{fig:coverage}, we compare the empirical coverage probability
\cite{Raftery2007,Gneiting2007} for different error models of $S_{1n}$. For each model, a posterior predictive distribution for the validation dataset is compared with the percentage of true validation data points that fall within a given credible interval. While each model performs well for this case, both heteroscedastic models outperform the homoscedastic model. To quantify the overall quality of uncertainty estimates across all credible intervals, we show the MACE \eqref{eq:mace} in Table~\ref{tab:coverage_comparison}. 
The HoEM generally exhibits the largest average spread about the mean, however, the optimal functional form of the error model depends on the observable under consideration. For example, for the single proton separation energy, the quadratic HeEM exceeds both the homoscedastic error model and the linear error model. However, for $S_{2p}$, it exhibits the worst performance of any error model across all observables. This demonstrates the importance of exploring different error models  when applying the BMC across multiple observables.

\begin{table}[ht!]
    \caption{Mean absolute calibration error (MACE) on the holdout validation set for proton and separation energies. Lower MACE is better.}
    \label{tab:coverage_comparison}
    \begin{ruledtabular}
\begin{tabular}{l cc c cc c c}
 & \multicolumn{2}{c}{Neutron} & & \multicolumn{2}{c}{Proton} & & \\
Model & $S_{1n}$ & $S_{2n}$ & & $S_{1p}$ & $S_{2p}$ & & Avg. \\
\hline\\[-7pt]
HoEM            & 6.07 & 3.21 & & 7.14 &  7.38 & & 5.95 \\
Linear HeEM     & 2.14 & 2.97 & & 5.52 &  6.42 & & 4.26 \\
Quadratic HeEM  & 3.21 & 2.38 & & 4.03 & 10.47 & & 5.02 \\
\end{tabular}
    \end{ruledtabular}
\end{table}

\begin{figure}[hbtp]
    \includegraphics[width=1\linewidth]{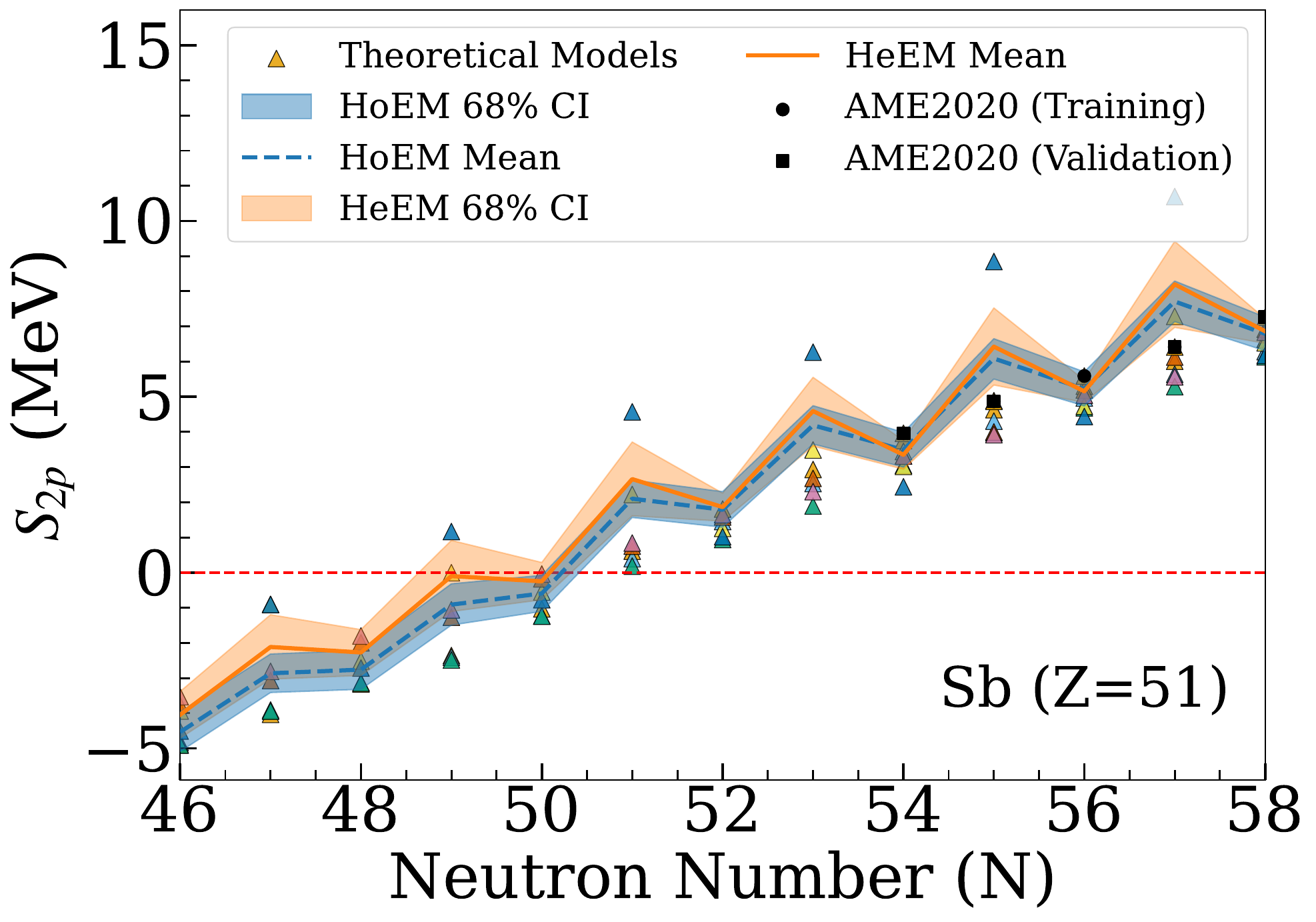}
    \caption{HoEM and linear HeEM predictions for $S_{2p}$ of the Sb ($Z=51$) chain alongside the  DFT model predictions. The HeEM band adapts to model divergence when $N$ is odd, while the HoEM uncertainty band remains constant across the entire chain.}
    \label{fig:s2p_sb}
\end{figure}

While the primary motivation for introducing a varying uncertainty scale is to account for the degradation in mean performance as the model extrapolates progressively farther from the training data, the choice of a varying error scale also enables the error model to reproduce other qualities of the theoretical models. 
Figure~\ref{fig:s2p_sb} demonstrates the qualitative difference between different approaches for a representative case of Sb~$(Z=51)$. The two error models produce similar central values and similar uncertainty bounds for even $N$. However, for odd-$N$ nuclei, model predictions strongly deviate from experiment, suggesting missing physics in theoretical description of odd-odd nuclei. Since the variance of model calculations $v_i$ is one of the predictors in Eq.~\eqref{Eq: sigma}, the HeEM BMC is naturally able to capture this behavior for odd-$N$ nuclei.

The farthest extrapolation from experimental data is for the drip lines, hence the limits of nuclear existence are uncertain. The most dramatic difference between our homoscedastic and heteroscedastic implementations manifests precisely at these limits. While the HoEM maintains an essentially constant uncertainty magnitude by construction, the HeEM reveals a progressive increase in predicted error, with uncertainties growing 5 keV on average per neutron number away from known data across all observables until $N = 107$. As the theoretical models diverge for $N \geq 108$, the predicted uncertainty increases dramatically (see Fig.~\ref{fig:syntheticdataexperiment}),  more than doubling on average in magnitude relative to the known data region.
This behavior highlights a key strength of the heteroscedastic approach: by linking the error scale directly to the inter-model spread $v_i$ rather than to a fixed function of distance alone, the HeEM is able to distinguish between regions where extrapolation is genuinely unconstrained and regions where, despite being far from data, the underlying physics still imposes a strong, shared constraint on the models.

The recent mass measurements of neutron-deficient Cd isotopes reported
in Ref.~\cite{lange2026constraining} provide an opportunity to test the performance of our BMC frameworks. As detailed in Table~\ref{tab:Cd_predictions}, the measurements for $S_{1n}$ at the closed shell demonstrate reasonable agreement with HeEM producing predictions within $1\sigma$ agreement for $^{96}$Cd and $^{98}$Cd. Moving just below the shell closure exposes the limitations of the constituent EDFs in capturing the extreme odd-even staggering in the empirical mass surface. At $^{97}$Cd, the experimental $S_{1n}$ drops precipitously, which is not captured by  our DFT models.  This common bias is not captured by the inter-model spread $v_i$ driving the HeEM error scale, and so the predicted uncertainty can underestimate the true deviation.

\begin{table}[ht!]
    \caption{Predictions and newly established experimental values (in MeV) for $S_{1n}$ in $^{96-98}$Cd. The experimental values are derived from the mass excesses reported in Ref.~\cite{lange2026constraining}, utilizing the extrapolated mass of $^{95}$Cd for the $^{96}$Cd calculation.}
    \label{tab:Cd_predictions}
    \begin{ruledtabular}
\begin{tabular}{l l ccc}
Obs. & Model & $^{96}$Cd & $^{97}$Cd & $^{98}$Cd \\
\hline\\[-7pt]
$S_{1n}$ & HoEM           & 16.61(41) & 14.15(41) & 15.42(42) \\
         & Linear HeEM    & 16.65(43) & 14.13(35) & 15.46(42) \\
         & Quadratic HeEM & 16.68(46) & 14.11(34) & 15.51(50) \\[4pt]
         & Exp.           & 16.99(14) & 12.91(4)  & 15.21(3)  \\
\end{tabular}
    \end{ruledtabular}
\end{table}

\begin{figure}
    \centering
    \includegraphics[width=1\linewidth]{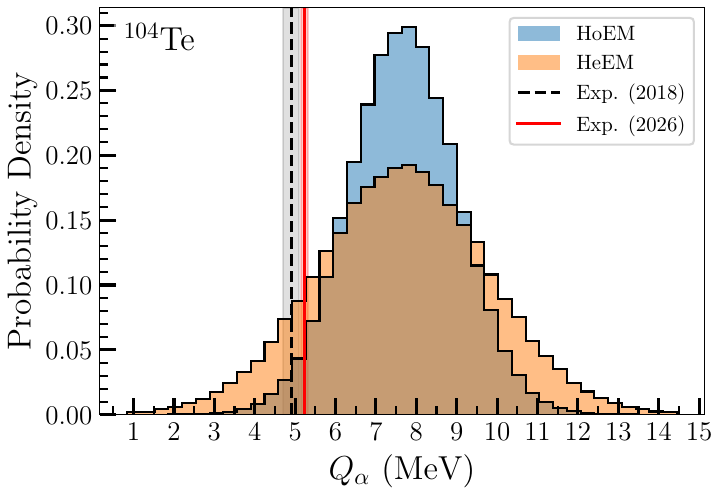}
    \caption{A histogram of all generated HoEM and linear HeEM predictions for $Q_\alpha$ for $^{104}$Te in the posterior alongside the old \cite{Auranen2018} and recent \cite{Cox2026} experimental values.}
    \label{fig:Qalpha}
\end{figure}

Prior to recent direct observations, the empirical properties of $^{104}$Te near the doubly magic $^{100}$Sn boundary were constrained by Ref.\,\cite{Auranen2018}, which reported a decay energy of 4.9(2) MeV.  The recent study~\cite{Cox2026}, isolated the  ground-state $\alpha$-decay chain to reveal a significantly higher decay energy of $Q_\alpha=5.23(9)$ MeV. As demonstrated in Fig.~\ref{fig:Qalpha}, the new BMC framework provides better agreement with both measurements than the previous HoEM framework.  However, similar to Ref.~\cite{lange2026constraining}, each of the individual DFT models predicts a higher $Q_\alpha$ value. This applies not just to $^{104}$Te, but also down the rest of the Te chain all the way to $^{112}$Te. This demonstrates how these $Q$ values can be used  as calibration data to further improve EDFs in this region.

To illustrate the practical impact of heteroscedastic uncertainties, we present detailed predictions for four neutron-deficient tin isotopes in Table \ref{tab:BMC_predictions_updated}. These nuclei ($^{96-99}$Sn) lie beyond current experimental reach on the proton-rich side and demonstrate how the HeEM and HoEM approaches yield similar mean predictions but  different uncertainty estimates. 
Even for the same nuclei, the outcome varies for different observables, suggesting that agreement between theoretical models is not necessarily uniform across observables.

\begin{table}[ht!]
    \caption{Predictions in MeV for $^{96-99}$Sn nuclear observables. The linear HeEM was used for this data.}
    \label{tab:BMC_predictions_updated}
    \begin{ruledtabular}
\begin{tabular}{l l cccc}
Obs. & Model & $^{96}$Sn & $^{97}$Sn & $^{98}$Sn & $^{99}$Sn \\
\hline\\[-7pt]
$BE$     & HoEM & 762.09(84)  & 777.96(84) & 795.10(86)  & 810.76(84) \\
         & HeEM & 762.17(101) & 778.03(94) & 795.15(101) & 810.79(88) \\[4pt]
$S_{1n}$ & HoEM & 17.87(28)   & 15.32(29)  & 17.14(29)   & 14.63(30)  \\
         & HeEM & 17.88(24)   & 15.30(24)  & 17.13(23)   & 14.59(28)  \\[4pt]
$S_{2n}$ & HoEM & 33.63(32)   & 33.09(34)  & 32.38(34)   & 31.72(34)  \\
         & HeEM & 33.66(32)   & 33.11(31)  & 32.35(34)   & 31.66(35)  \\[4pt]
$S_{1p}$ & HoEM &   0.84(37)  &  1.44(39)  &  1.96(38)   &  2.46(38)  \\
         & HeEM &   0.77(32)  &  1.37(32)  &  1.89(31)   &  2.39(33)  \\[4pt]
$S_{2p}$ & HoEM & $-$0.44(44) &  0.66(43)  &  1.90(44)   &  2.93(43)  \\
         & HeEM & $-$0.61(53) &  0.47(50)  &  1.75(47)   &  2.85(37)  \\
\end{tabular}
    \end{ruledtabular}
\end{table}

\begin{figure*}[!htbp]
    \centering
    \includegraphics[width=1\linewidth]{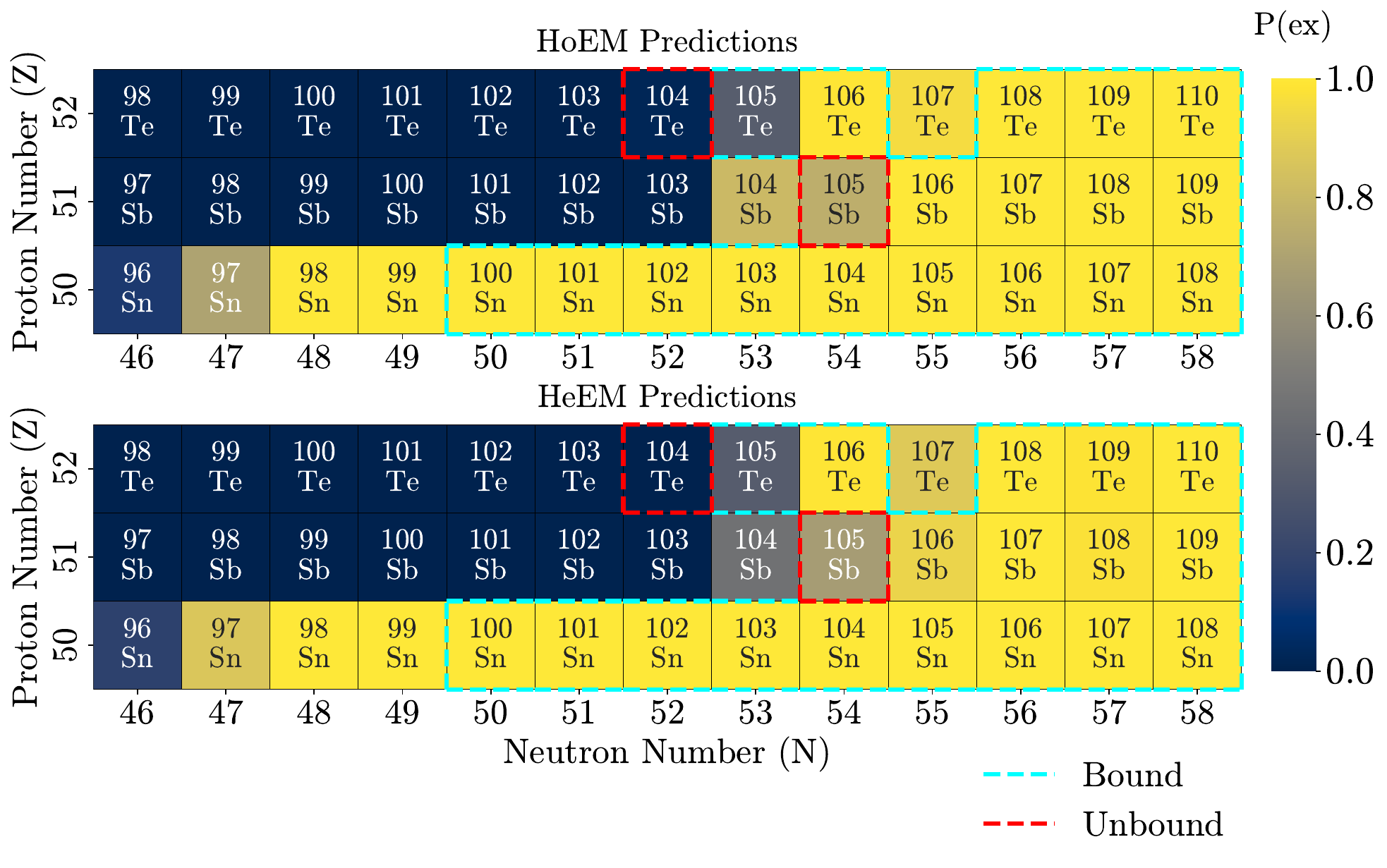}
    \caption{A heatmap of the probability of existence  (\ref{eq:Pex}) for bound and unbound nuclei predicted using homoscedastic and heteroscedastic approaches. The linear HeEM was used for the heteroscedastic predictions.}
    \label{fig:heatmap}
\end{figure*}

The proton-rich frontier presents a challenging test case for local extrapolation. Figure \ref{fig:heatmap} displays the predicted probabilities of existence for nuclei in the tin region approaching the proton drip line. The probability of existence \cite{Neufcourt2020b} is defined as the probability that the selected nucleus is bound to one- and two-particle emission:
\begin{equation} P(\text{ex})(Z,N) = P\big(S_{1{\mathfrak n}}(Z,N) > 0,\ S_{2{\mathfrak n}}(Z,N) > 0\big), \label{eq:Pex}
\end{equation} where ${\mathfrak n} \in \{n,p\}$ denotes the neutron or proton rich region, and the probability is computed over the posterior predictive distribution of the BMC model. 

While both error models largely agree on the location of the drip line, they differ substantially in their predictions for individual nuclei in this region. For example, the transition gradients from confidently bound to unbound exhibit distinct structural differences between the two frameworks. The HoEM, constrained by a static error scale, predicts a relatively sharp transition across the isotonic lines, with probabilistic assignments for transitional nuclei that remain rigid. In contrast, the HeEM produces a probability gradient that correlates with local model divergence. While the HoEM may confidently classify a transitional nucleus as unbound, the expanded uncertainty bands of the HeEM at the extrapolative frontier allow for a non-trivial probability of existence. A negative central separation-energy prediction does not, by itself, imply that a nucleus is theoretically excluded from existing, and only a fully probabilistic treatment of the uncertainty can reveal whether such marginal cases remain experimentally viable targets. Both the HoEM and the HeEM confidently report that $^{104}$Te is proton unbound, a claim that agrees with the AME2020 record of the two proton separation energy as $-0.73$ MeV \cite{AME2020}.

\begin{figure}
    \centering
    \includegraphics[width=1\linewidth]{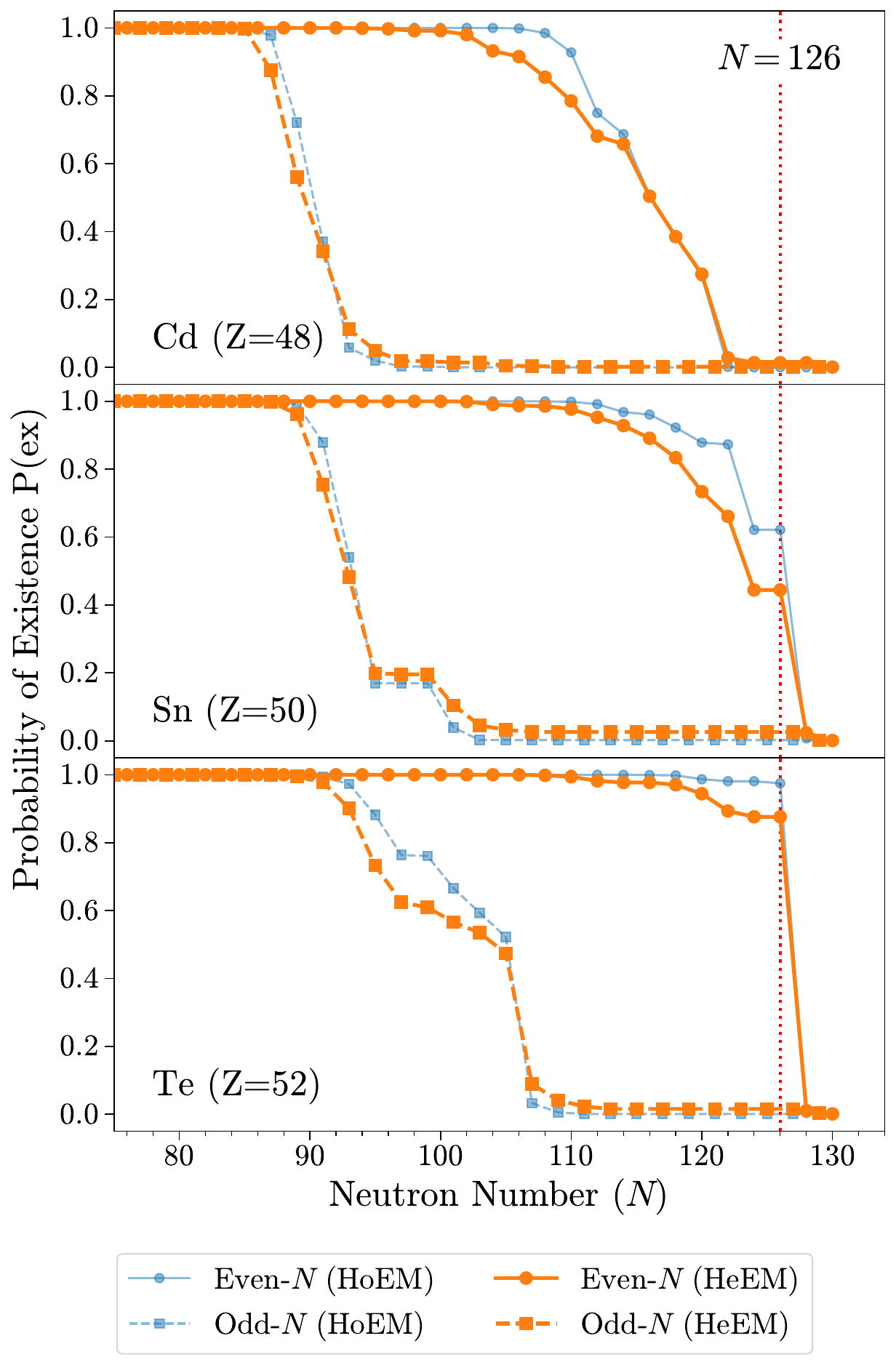}
    \caption{Probability of existence for even-$Z$ chains in the tin region. Predictions are made using the linear HeEM.}
    \label{fig:neutrondrip lines}
\end{figure}

The same process was used to predict neutron drip lines in Fig.~\ref{fig:neutrondrip lines}. Similar to the proton drip line, the two error models identify similar locations for the dripline. However, for all three chains, the linear HeEM produces a gentler dropoff in the likelihood of existence than the HoEM, reflecting the genuine theoretical disagreement as to where exactly each chain becomes unbound, rather than the artificially sharp cutoff imposed by a fixed error scale. Across all models and chains, there is locally boosted model consensus at the $N=126$ shell gap, where both the HoEM and HeEM predict a rapid decrease in existence likelihood. 

\section{Summary and Conclusions}
\label{sec:conc}

In this work, we presented a novel approach for nuclear uncertainty quantification by introducing physically-motivated heteroscedasticity into the BMC framework of nuclear DFT models. While previous benchmarking of nuclear mass models has largely relied on an assumed constancy of the variance of predictions across the nuclear landscape, our results demonstrate that this simplification fails to account for the inherent degradation of model predictive power as one extrapolates toward the limits of nuclear stability. By allowing the error scale to vary across isotopes through a dependency on the disagreement across models and the principal component distance, the HeEM provides a more robust and statistically rigorous description of the systematic uncertainties associated with the nuclear mass surface.

Before applying this framework to experimental data, we validated it using a synthetic data test, in which each constituent DFT model was treated as a surrogate for the experimental data. This procedure allowed the simulation of genuine extrapolation to confirm that the HeEM consistently produces better-calibrated uncertainties than the HoEM and to motivate our choice of the linear HeEM as the primary model for the AME2020 application.

Using HeEM, we were able to achieve strong performance metrics (RMSE, $\chi^2$, and MACE) on the validation set for AME2020 for every observable we investigated, whereas the HoEM was either conservative or overconfident. We identified the mean absolute calibration error as a strong predictor of extrapolative performance. Specifically, models that perform well on the validation set at the boundaries of known data, such as the linear HeEM, maintain favorable performance well into the predictive domain. In the extreme extrapolation regime for the neutron rich region ($N\geq 100$), we observed a sharp surge in uncertainty, marking the transition from well-constrained theoretical predictions to domains where model agreement deteriorates.

The practical impact of this framework is most evident in the prediction of the nuclear drip lines. While homoscedastic and heteroscedastic approaches may yield similar mean predictions, their probabilistic assessments of nuclear existence differ substantially in critical regions. 
Ultimately, this study underscores the necessity of moving beyond global error scales in nuclear theory. By adopting heteroscedastic uncertainties that correlate with local model divergence and distance from known data, we provide a more faithful representation of the limits of our theoretical understanding.

\section*{Acknowledgments}
Discussions with Franziska Maier are gratefully acknowledged. This material is based upon work supported by the U.S. Department of Energy under Award Numbers DE-SC0013365 and DE-SC0023688 (Office of Science, Office of Nuclear Physics), DOE-DE-NA0004245 (NNSA, the Stewardship Science Academic Alliances program), DE-SC0023175 (Office of Science, NUCLEI SciDAC-5 collaboration), and by the National Science Foundation under award number 2004601 (CSSI program, BAND collaboration).


\providecommand{\noopsort}[1]{}\providecommand{\singleletter}[1]{#1}%

\end{document}